# Spin current driven by ultrafast magnetization of FeRh

Kyuhwe Kang[1], Hiroki Omura[2], Oukjae Lee[3], Kyung-Jin Lee[4], Hyun-Woo Lee[5], Tomoyasu Taniyama[2], Gyung-Min Choi[1,6*]

[1]Department of Energy Science, Sungkyunkwan University, Suwon 16419, Korea

[2]Department of Physics, Nagoya University, Nagoya 464-8601, Japan

[3]Center for Spintronics, Korea Institute of Science and Technology, Seoul 02972, Korea

[4]Department of Physics, Korea Advanced Institute of Science and Technology, Daejeon 34141, Korea

[5]Department of Physics, Pohang University of Science and Technology, Pohang 37673, Korea

[6]Center for Integrated Nanostructure Physics, Institute for Basic Science, Suwon 16419, Korea

Laser-induced ultrafast demagnetization is an important phenomenon that probes arguably ultimate limits of the angular momentum dynamics in solid. Unfortunately, many aspects of the dynamics remain unclear except that the demagnetization transfers the angular momentum eventually to the lattice. In particular, roles of electron-carried spin current are debated. Here we experimentally probe the spin current in the opposite phenomenon, i.e., laser-induced ultrafast magnetization of FeRh, where the laser pump pulse initiates the angular momentum build-up rather than its dissipation. Using the time-resolved magneto-optical Kerr effect, we directly measure the ultrafast-magnetization-driven spin current in a FeRh/Cu heterostructure. Strong correlation between the spin current and the net magnetization change rate of FeRh is found even though the spin filter effect is negligible in this opposite process. This result implies that the angular momentum build-up is achieved by an angular momentum transfer from the electron bath (supplier) to the magnon bath (receiver) and followed by the spatial transport of angular momentum (spin current) and dissipation of angular momentum to the phonon bath (spin relaxation).

**Angular momentum transfer during ultrafast demagnetization**

Ultrafast demagnetization is a rapid quenching of the magnetic ordering of a ferromagnetic metal (FM) in less than a few picoseconds[1-4]. Such a short timescale indicates a rapid dissipation mechanism for angular momentum in FM. Considering the Einstein-de Haas effect, the ultimate destination of angular momentum should be the lattice bath[5-7]. A recent study in a single FM layer demonstrated that ultrafast demagnetization induces a circularly polarized phonon in less than a picosecond, suggesting a fast transfer of angular momentum between the magnetization bath and phonon bath[8]. For this fast dissipation of angular momentum to the phonon and lattice in a single FM, the role of the conduction electrons is hidden.

The electronic contribution to the angular momentum dissipation was revealed from the spin dynamics in heterostructures: ultrafast demagnetization of FM generates a transient spin current in a non-magnetic metal (NM). Ultrafast-demagnetization-driven spin currents have been confirmed by various experimental observations, such as spin accumulation on NM in FM/NM[9–13], terahertz generation from NM in FM/NM[14-17], coupling of demagnetization dynamics of FM in FM/NM/FM with a collinear magnetization of the FM pair[18-20], and spin-transfer-torque on FM in FM/NM/FM with a non-collinear magnetization of the FM pair[21,22]. However, a theoretical interpretation of the spin current remains controversial[9,10,23-25]. The superdiffusive theory argues that the spin-dependent transport of hot electrons inside FM generates a strong spin current to NM[23,24]. Since the electronic density-of-states of FM is spin-dependent, an electronic transport at the FM/NM interface leads to a spin filter effect[26], and the superdiffusive theory is a hot-electron version of the spin filter effect. Another possible mechanism is the angular momentum transfer between magnons (wave-like excitation of local magnetic moments) and conduction electrons[4,10,27,28]. The angular momentum of the FM phase is stored in the $d$-band electrons, which are responsible for magnetization, and angular momentum transfer should be mediated by the spin excitations of the $d$-band, such as magnons[27,28] and Stoner excitations[4] (in this work, for simplicity, we call magnons as a representative of the spin excitations). When the angular momentum loss in the magnon bath ($d$ band) is converted to the angular momentum gain in the electron bath ($sp$ band), a diffusive spin current is generated across the FM/NM interface[10].

In this study, we investigated the spin current in the reverse process, i.e., ultrafast magnetization of FeRh during the phase transition from the antiferromagnetic metal (AFM) phase to the FM phase. Whereas ultrafast demagnetization dissipates angular momentum, ultrafast magnetization of FeRh should absorb angular momentum from environment. Since the spin filter effect is not allowed with the initial AFM phase, ultrafast magnetization is an optimal circumstance to investigate the mechanism for the spin current generation.

**Angular momentum transfer during phase transition**

Angular momentum transfer is also important to understand the mechanism of the phase transition. FeRh exhibits a unique phenomenon of the 1$^{st}$-order phase transition, which leads to a magnetic transition from the AFM phase to the FM phase accompanied by a lattice expansion of ~1%, at a critical temperature of approximately 350 K[29,30]. Previous reports have demonstrated that the magnetization change and lattice expansion can occur on the order of picoseconds during the phase transition[31–37]. However, the mechanism for such a fast phase transition is under debate. The timescale of the phase transition consists of two contributions: the timescale for the driving force and the timescale for angular momentum transfer. As the driving force, Kittel proposed that drastic lattice expansion drives the sign inversion of the exchange energy[38], then its timescale could be related to the speed of lattice expansion[35,37]. Note that other mechanisms of electronic band structure[39], magnetic moment of Rh atom[32], and magnon excitations[40] had been proposed as well for the driving force, whose timescale could be different from the speed of lattice expansion. Once the driving force is applied, considering the huge difference in the magnetization density between the AFM and FM phases, a rapid transfer of angular momentum should occur during the phase transition. However, the exact procedure of the angular momentum transfer during the phase transition is not known. Especially, the role of conduction electrons for the angular momentum transfer has not been studied.

In this study, we investigate angular momentum transfer during the phase transition of FeRh by measuring the ultrafast-magnetization-driven spin current in the MgO substrate/FeRh (20 nm)/Cu (120 nm) structure. The FeRh and Cu layers were grown by Molecular Beam Epitaxy (see Methods), and the FeRh/Cu interface has a clean and flat morphology (Fig. 1a). From the quasi-static measurements of the magnetization as a function of temperature, the phase of our FeRh changes from AFM to FM at ~370 K (Fig. 1b and Supplementary Section 1). The transient dynamics of the phase transition is investigated by using an optical pump-probe technique (see Methods and Supplementary Section 2). When we inject a pump pulse on FeRh through the MgO substrate, a local heating triggers the phase transition of FeRh on a timescale of a few picoseconds. A probe pulse detects the magnetization ($\Delta M_{FeRh}$) of FeRh and spin accumulation ($\Delta M_{Cu}$) of Cu via magneto-optical Kerr effect, and it also detects the lattice expansion ($\Delta L$) of FeRh via the strain-induced reflectivity change (Fig. 1c). We examine the correlation between the magnetization dynamics of FeRh and the spin accumulation of Cu in terms of sign, time delay, and magnitude.

**Ultrafast magnetization of FeRh**

Firstly, we measure ultrafast magnetization of FeRh during the phase transition. With an initial AFM phase at a base temperature of 300 K, a pump pulse triggers an ultrafast phase transition from the AFM phase to the FM phase, and a probe pulse detects the evolution of net magnetization ($\Delta M_{FeRh}$), thereby probing the magnetic transition in a time-resolved manner. With a pump fluence of >2 J m$^{-2}$, a sharp increase of $\Delta M_{FeRh}$ occurs within 4 ps (Fig. 2a). However, with a pump fluence of 0.9 J m$^{-2}$, $\Delta M_{FeRh}$ becomes negligible. Such a threshold is a characteristic behavior for the 1$^{st}$ order phase transition with a latent heat. The magnitude of $\Delta M_{FeRh}$ at 4 ps saturates with a pump fluence of 7 J m$^{-2}$, whereas $\Delta M_{FeRh}$ at a longer timescale of ~100 ps exhibits further increase with increasing the pump fluence (Fig. 2b). Both fast timescales of a few ps and slow timescales of ~100 ps for the FeRh phase transition have been observed previously, and the former and latter were attributed to the FM domain nucleation and expansion, respectively[33-35]. In this study, we focus on the timescale of 0–10 ps because only the fast dynamics plays a significant role in the spin current generation.

To emphasize the timescale of the ultrafast phase transition, we compare the phase-transition-driven dynamics to ultrafast-demagnetization-driven one. Whereas the phase transition induces a net magnetization starting from zero, ultrafast demagnetization decreases magnetization starting from a finite value. Upon increasing the base temperature to 430 K, the initial phase of FeRh became a FM phase. A sudden heating of the FM phase by a pump pulse leads to ultrafast demagnetization within 1 ps (Fig. 2c). The timescale of ultrafast demagnetization corresponds to the thermalization time among the electron, magnon, and phonon baths (Supplementary Sections 3 and 4). Importantly, when we compare the decrease of $\Delta M_{FeRh}$ by ultrafast demagnetization and the rise of $\Delta M_{FeRh}$ by ultrafast magnetization, the timescale of the ultrafast magnetization is delayed by 2.5 ps. Such a delay suggests that the phase transition requires an additional process other than the thermalization. From the reflectivity change measurement (Supplementary Section 5), we observe the same 2.5 ps delay in the lattice expansion. When the lattice expansion starts near the FeRh surface, it propagates through the FeRh layer as a strain wave. Then, the 2.5 ps delay can be explained as $(d_{\text{FeRh}} - d_{\text{surf}})/v_s$, where $d_{\text{FeRh}}$ is the thickness of FeRh of 20 nm, $d_{\text{surf}}$ is the surface depth of initial heating, and $v_s$ is the sound velocity of ~5 km s$^{-1}$ of FeRh[41]. This result indicates that the timescale of the lattice expansion is the limiting factor for the speed of the phase transition, and the timescale of the angular momentum transfer is sufficiently fast not to be a limiting factor.

**Spin accumulation on Cu**

Next, we measure the spin accumulation on Cu driven by ultrafast magnetization of FeRh in the FeRh/Cu heterostructure. While a pump pulse triggers the phase transition of FeRh, a probe pulse detects spin accumulation in the conduction electron bath of Cu ($\Delta M_{\text{Cu}}$). For NM materials that have no magnon bath, only the magnetization of the conduction electrons is responsible for the Kerr rotation. We observe a negative spin polarization on Cu with a peak position at 4 ps, clearly indicating the electron-carried spin current from FeRh to Cu (Fig. 3a). For comparison, we also measure ultrafast-demagnetization-driven spin accumulation with an initial FM phase of FeRh at a base temperature of 430 K. Ultrafast demagnetization generates a positive spin polarization on Cu with a peak position at 1 ps, which is 3 ps faster than that from the phase transition.

The sign and time delay of the spin accumulation on Cu provide an important clue for the underlying mechanism. The opposite sign of $\Delta M_{\text{Cu}}$ driven by the ultrafast magnetization and demagnetization of FeRh disproves any mechanism based on the spin-filter effect. The sign of the spin polarization by the spin filter effect is determined by the magnetization direction of the FM phase of FeRh, which is set by an external magnetic field and it is the same for ultrafast magnetization and demagnetization. In addition, the 4 ps time delay in $\Delta M_{\text{Cu}}$ cannot be explained by hot electron effect because the transport of hot electrons takes less than 1 ps (Supplementary Section 3). On the other hand, the sign and time delay in $\Delta M_{\text{Cu}}$ has a close relation to the magnetization dynamics of FeRh. Assuming that the magnetization loss in the magnon bath is converted to the spin generation in the electron bath, we compare $\Delta M_{\text{Cu}}$ to the time-derivative of magnetization of magnon bath, $-dM_{\text{FeRh}}/dt$, in Fig. 3b. The sign of $\Delta M_{\text{Cu}}$ matches well to the sign of $-dM_{\text{FeRh}}/dt$: ultrafast magnetization/demagnetization of FeRh induces a negative/positive $\Delta M_{\text{Cu}}$. In addition, a time delay of 3 ps in $\Delta M_{\text{Cu}}$ is close to that of 2.5 ps between the ultrafast magnetization and demagnetization in $-dM_{\text{FeRh}}/dt$. This result reveals a critical role of the angular momentum transfer between the magnon and electron baths for the spin current generation.

For a quantitative analysis of the angular momentum transfer, we perform a simulation of the combined process of spin generation, diffusion, and relaxation in the conduction electrons (see Methods and Supplementary section 6). For the spin generation, we assume that magnon-electron coupling is much stronger than magnon-phonon coupling in FeRh so that angular momentum transfer is always mediated by electrons. This assumption is often used for analysis of ultrafast demagnetization of FM[1,4] and is supported by a slow demagnetization in insulating ferrimagnets without the conduction electron bath[42]. Then, $\Delta M_{FeRh}$ of the magnon bath is entirely converted to the spin polarization of the conduction electron bath, and the spin generation rate ($g_s$) of the electron bath is expressed as $g_s = -dM_{FeRh}/dt$. The generated spins in the electron bath can diffuse from FeRh to Cu *via* spin diffusion and dissipate to phonons *via* spin relaxation. The simulation result for the spin diffusion well explains the peak position of $\Delta M_{Cu}$: the negative peak at 4 ps for the phase-transition-driven $\Delta M_{Cu}$ is the combined result of the time scale for the phase transition of FeRh ($\Delta t_p$ = 2.5 ps) and that for the diffusive transport from FeRh to Cu ($\Delta t_d$ = 1.5 ps) (Fig. 3c), and the positive peak at 1 ps for the ultrafast-demagnetization-driven $\Delta M_{Cu}$ is almost exclusively determined by $\Delta t_d$ (Fig. 3d). Note that $\Delta t_d$ is approximately 0.5 ps longer in Fig. 3c than in Fig. 3d because the electron mobility of FeRh is smaller in the AFM phase than in the FM phase.

From the magnitude of $\Delta M_{Cu}$, we determine the spin relaxation time ($\tau_s$) of FeRh, which describes the speed of angular momentum dissipation from the electron bath to the phonon bath. The longer $\tau_s$ of FeRh, the larger the amount of spin of electrons can diffuse from FeRh to Cu before the dissipation to phonons (the long $\tau_s$ of Cu has a negligible effect on the simulation result). Fitting the amplitude of the spin accumulation between experiment and simulation, we determine $\tau_s$ of 0.05 ps for the AFM phase of FeRh. We also perform a similar simulation for the ultrafast-demagnetization-driven spin accumulation and determine $\tau_s$ of 0.14 ps for the FM phase of FeRh. The difference in $\tau_s$ between the AFM and FM phases could be due to the different band structures of electrons, magnons, and phonons. The $\tau_s$ can be converted to the spin diffusion length, $l_s$, using the relation of $l_s = \sqrt{D\tau_s}$, where $D$ is the electronic diffusivity of FeRh, and we obtain $l_s$ values of 2.1 nm and 2.9 nm, respectively, for AFM and FM phases of FeRh, which are smaller than $l_s$ of 7 nm of pure Fe[43]. According to the Elliot-Yafet mechanism[4,44,45], the spin relaxation of the electron bath is due to incoherent electron-phonon scattering in the presence of spin-orbit coupling. We expect that the strong spin-orbit coupling of Rh[46] enhances the spin-flip probability during the electron-phonon scattering.

We propose a process for angular momentum transfer during the phase transition of FeRh (Fig. 4). When electrons are excited by pump light, rapid energy transfer occurs from the electron bath to the magnon and phonon baths. This thermalization is responsible for the timescale of ultrafast demagnetization of the FM phase. For the phase transition from the AFM phase to FM phase, an additional timescale for the lattice expansion is required. For the angular momentum transfer during the phase transition, the conduction electron bath quickly supplies the required angular momentum for the magnon bath. Then, a positive $\Delta M$ of the magnon bath leads to a negative $\Delta M$ of the conduction electron bath of FeRh. The angular momentum of the conduction electron should relax to the phonon bath, and subsequently to the lattice. However, before a complete relaxation, a portion of angular momentum in the conduction electron can spatially diffuse, leading to transient spin current to an adjacent layer of Cu. The magnitude of spin current is determined by the competition between the diffusion time and relaxation time. The same process should be responsible for angular momentum transfer during ultrafast demagnetization of FM. The only differences are the time scale (no lattice expansion) and the sign of spin current. The electron enables the rapid transfer of angular momentum during the ultrafast change of magnetization, either magnetization or demagnetization, because of the strong electron-magnon and electron-phonon coupling.

Our work combines two separate phenomena: spin current and phase transition. An integrated understanding of the spatial flow of angular momentum (spin current) and the relocation of angular momentum inside a material (phase transition) will expand the research area of spintronics. In addition, a dynamic coupling between the spin current and phase transition could be useful for the high-speed operation of memory devices, such as magnetic memory and phase change memory.

**Methods**

*Film growth*

A MgO (001) substrate/Fe$_{50}$Rh$_{50}$ (20 nm)/Cu (120 nm)/SiO$_2$ (4 nm) stacking structure was fabricated using molecular beam epitaxy (MBE) and sputtering. An epitaxial FeRh layer was grown on a MgO (001) substrate at 450 °C by co-evaporating Fe and Rh from separate sources in an ultrahigh vacuum MBE chamber with a base pressure of ~ 10$^{-10}$ Torr, followed by post-annealing at 600 °C. To achieve a strong spin current, the thickness of FeRh should be close to the spin diffusion length of FeRh[43]. However, we found that the AFM phase of FeRh becomes incomplete when its thickness becomes too thin. The thickness of FeRh was chosen to 20 nm to have a complete AFM phase at 300 K (Supplementary Section 1). A 120-nm-thick Cu layer was then grown on the FeRh layer at room temperature. The thickness of Cu should be much thicker than the light penetration depth but smaller than the spin diffusion length of Cu[22]. At the Cu thickness of 120 nm, the probe on the Cu side only sees the spin accumulation on Cu without any contribution from the magnetization of FeRh. After the growth of FeRh and Cu, the films are immediately transferred to the sputter chamber, and an additional capping layer of a 4-nm-thick $SiO_2$ was grown on top of the Cu layer using RF sputter at Ar pressure of $2\times10^{-2}$ Torr at room temperature. The $SiO_2$ capping layer prevents the oxidation of Cu so that our films maintain its property.

*Optical measurement*

We used a pump-probe optical technique to observe both the phase transition of FeRh and the spin accumulation on Cu in the time domain. A pulsed laser was generated using a Ti-sapphire oscillator with a repetition rate of 80 MHz and a wavelength of 785 nm. The laser beam was divided into pump and probe beams by a polarizing beam splitter. The time delay between the pump and probe beams was controlled using a motorized mechanical stage. The pump and probe beams were modulated by an electro-optic modulator and optical chopper, respectively, at frequencies of 1 MHz and 200 Hz. The full widths at half maximum of the time correlation of the pump and probe were determined to be 1.2 ps. Considering the large group velocity dispersion of EOM, we expect that FWHM is 1.0 and 0.2 ps for pump and probe, respectively. Both the pump and probe beams were focused to a spot size of 3 μm ($1/e^2$ radius) using a 20X objective lens. The pump triggers the phase transition of FeRh, and the probe detects magnetization of FeRh or Cu via magneto-optical Kerr effect (MOKE) and lattice expansion of FeRh via strain-induced reflectivity change. For the MOKE, we used a polar MOKE geometry and aligned the magnetization of FeRh to the out-of-plain direction using a ring magnet. Depending on the position of probe, either on the FeRh side or on the Cu side, we used different optical setup, whose schematics and explanation are shown in Supplementary Section 2.

*Spin diffusion simulation*

To simulate the diffusive transport of the spins of conduction electrons, we used the following equation:

$$\frac{\partial \mu_s}{\partial t} = D \frac{\partial^2 \mu_s}{\partial z^2} - \frac{\mu_s}{\tau_s} + \frac{g_s}{\mu_B N_s},$$

where $\mu_s$ is the spin chemical potential, $t$ is time, $z$ is the spatial coordinate along the film thickness, $D$ is the diffusion constant, $\tau_s$ is the spin relaxation time, $g_s$ is the spin generation rate, $\mu_B$ is the Bohr magneton, and $N_s$ is the spin density of state. $D$ was determined from the electrical conductivity as $D = \frac{\sigma_e}{e^2 N_F}$, where $\sigma_e$ is the electronic conductivity of the material and $N_F$ is the electronic density of states at the Fermi level. $N_s$ was determined as $N_s = \frac{N_F}{2}$. $\mu_s$ of FeRh and Cu are connected at the interface with an interfacial spin conductance of $G_s$, which is determined as $G_s = \frac{G_e}{2e^2}$, where $G_e$ is the electrical conductance at the interface and $e$ is the elementary charge. $g_s$ is the spin generation rate on the electrons of FeRh, and it is obtained from the magnetization dynamics as $g_s = -\frac{dM}{dt}$ (there is no $g_s$ on the electrons of Cu). Further explanation on these parameters is described in Supplementary Section 6.

**Acknowledgments**

K.K. and and G.-M.C. are supported by Samsung Research Funding & Incubation Center of Samsung Electronics under Project Number SRFC-MA2001-03. H.O. and T.T. are supported in part by JST CREST Grant No. JPMJCR18J1, JSPS KAKENHI Grant No. 21H04614.

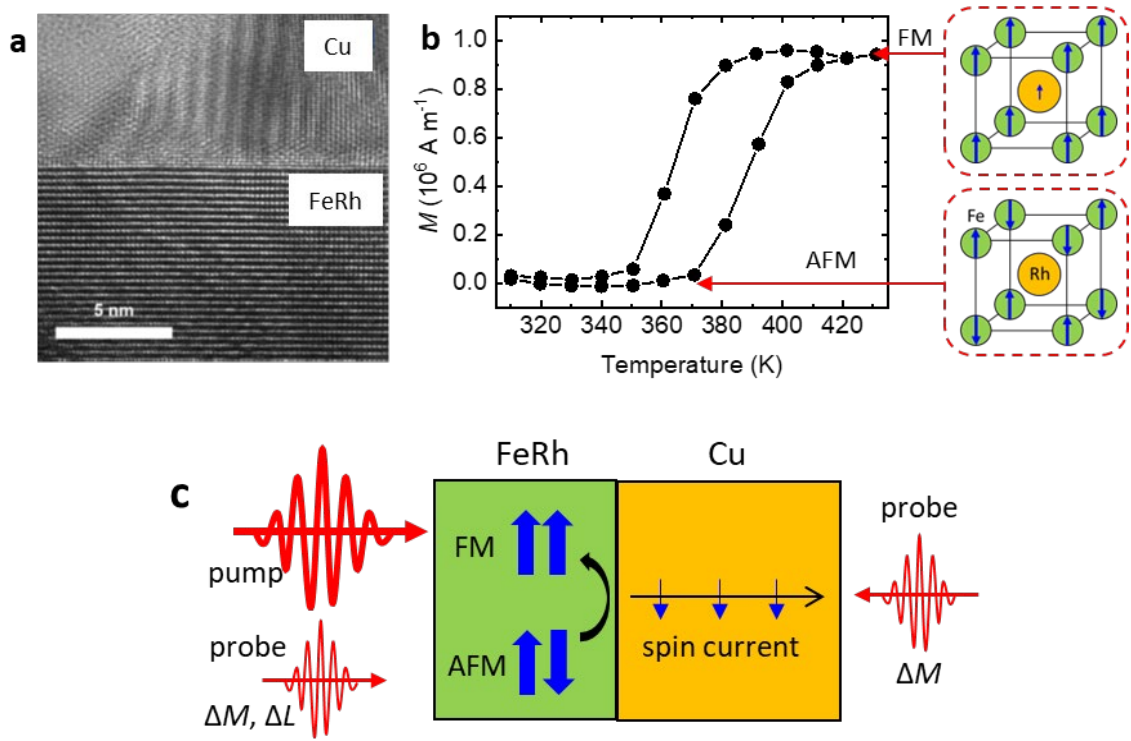

**Figure 1. Schematics of the experiment. a**, Transmission-electron-microscope image of the FeRh/Cu heterostructure. A white scale bar indicates 5 nm. The FeRh/Cu interface shows a clean and flat morphology. **b**, Magnetization versus temperature of FeRh. At temperature less than 350°C, FeRh becomes an AFM phase with negligible magnetization. At temperature above 400°C, FeRh becomes a FM phase with a net magnetization of ~$10^6$ A m$^{-1}$. The left insets are the configurations of atomic moments (blue arrows) of the AFM and FM phases of FeRh. **c**, Schematics of the ultrafast-magnetization-driven spin current. The pump pulse triggers the phase transition from the AFM phase to FM phase of FeRh, which induced a spin current to Cu, where the blue arrow indicates the spin polarization, the black arrow indicates the flow of spin. The probe pulse detects the magnetization ($\Delta M$) and lattice expansion ($\Delta L$) of FeRh and spin accumulation ($\Delta M$) of Cu.

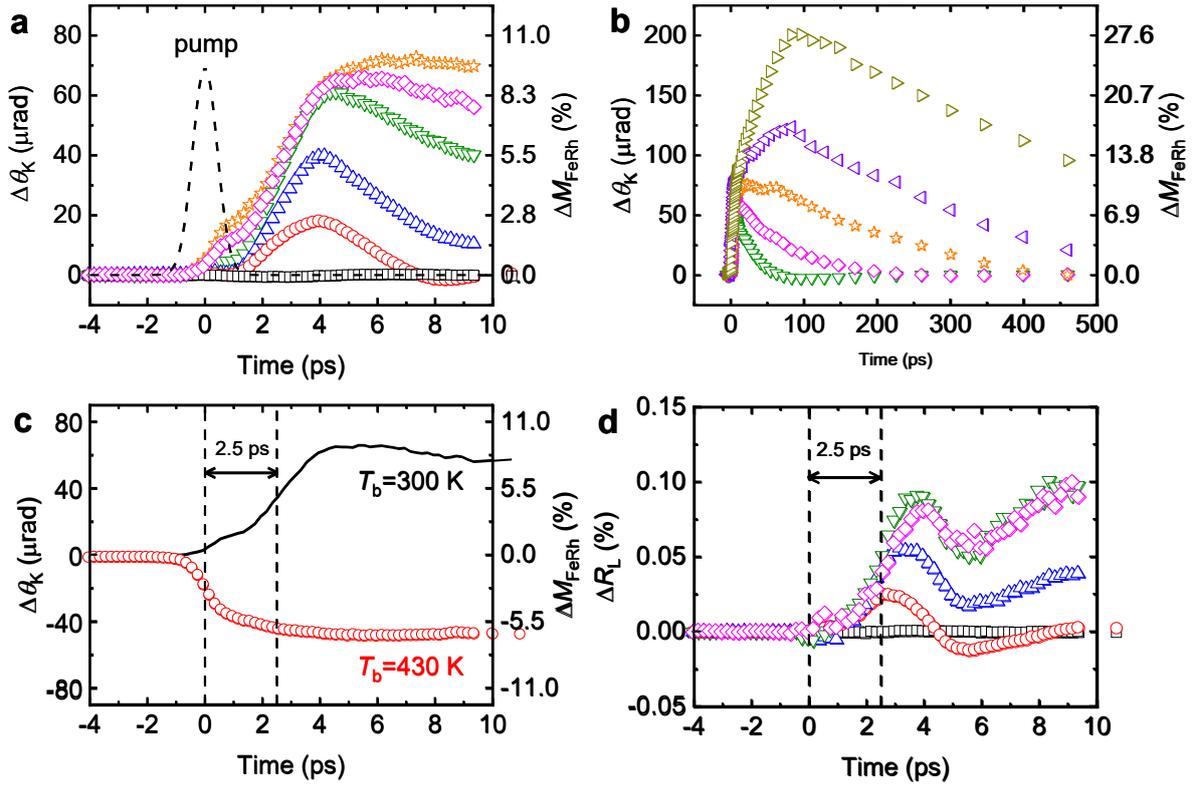

**Figure 2. Ultrafast magnetization of FeRh in FeRh/Cu heterostructure. a,b**, The dynamic Kerr rotation ($\Delta\theta_K$) by ultrafast magnetization of FeRh at (a) short timescale of 10 ps and (b) long timescale of 450 ps. The magnetization direction of the FM phase of FeRh is set by an external magnetic field of ±0.15 T, which is along the out-of-plane direction. The left y-axis is the dynamic Kerr rotation, and the right y-axis is the relative magnetization ($\Delta M_{FeRh}$). $\Delta M_{FeRh}$ is determined as $\Delta M = \frac{\Delta\theta_K}{\theta_K} \div \frac{M_z}{M_s}$, where $\theta_K$ is the static Kerr rotation of 5.8 mrad for the saturation magnetization ($M_s$) of the FM phase of FeRh, and $M_z$ is the z-component of magnetization. With an external field of 0.15 T, $M_z/M_s$ is 0.13. The color indicates the pump fluence in a unit of J m$^{-2}$: 0.9 (black square), 2.2 (red circles), 3.5 (blue up-triangles), 5.3 (green down-triangles), 7.1 (magna diamonds), 8.8 (orange stars), 10.6 (purple left-triangles), and 16.4 (dark yellow right-triangles). A dashed line at the time zero indicate the position of the pump pulse. **c**, The dynamic Kerr rotation by the ultrafast demagnetization of the FM phase of FeRh with a pump fluence of 7.1 J m$^{-2}$ at a base temperature ($T_b$) of 430 K (red circles). The black solid line is ultrafast magnetization of (a) at a base temperature of 300 K with the same pump fluence. There is 2.5 ps time delay between the ultrafast demagnetization and ultrafast magnetization. **d**, The lattice-expansion-driven reflectivity change ($\Delta R_L$) during the phase transition. The rise of $\Delta R_L$ occurs 2.5 ps after the pump pulse. The color in (d) indicates the pump fluence as in (a).

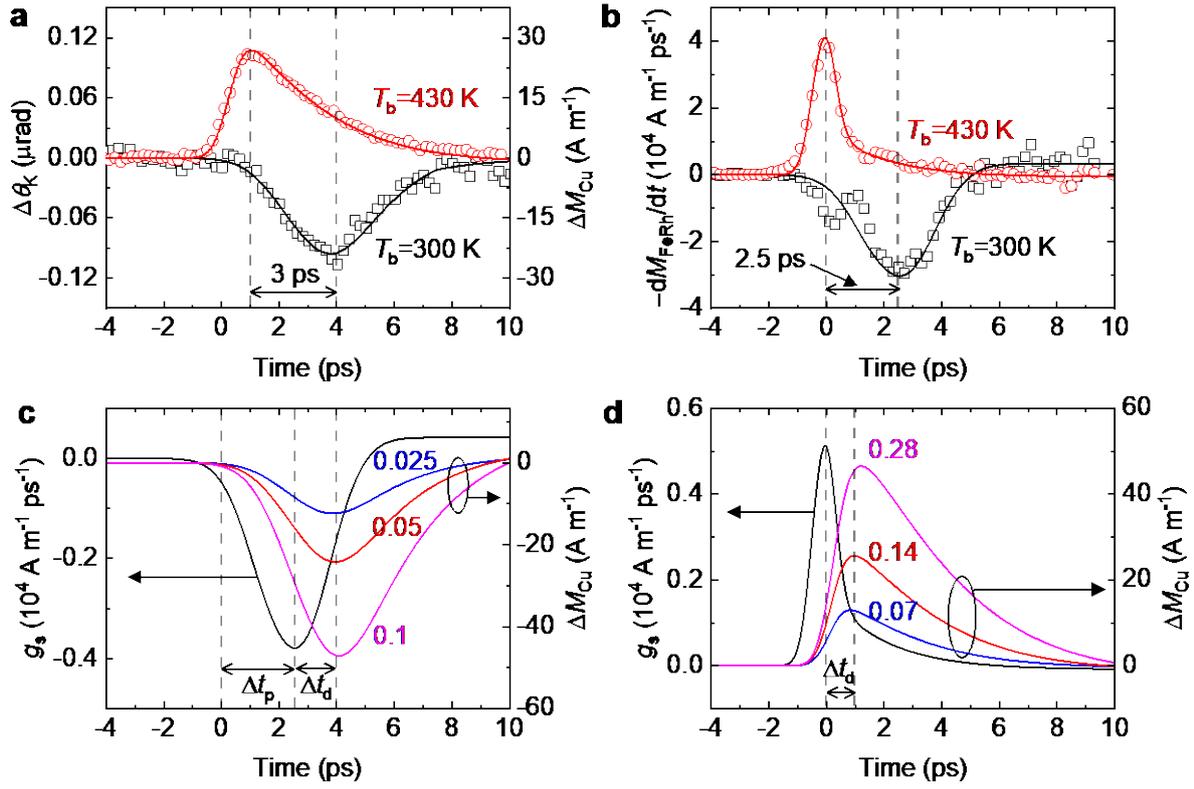

**Figure 3. Spin accumulation on Cu in FeRh/Cu heterostructure. a**, The dynamic Kerr rotation ($\Delta\theta_K$) by the spin accumulation on Cu driven by ultrafast magnetization (black squares) and ultrafast demagnetization (red circles) of FeRh, respectively, at a base temperature ($T_b$) of 300 K and 430 K. The pump fluence is fixed to 7.1 J m$^{-2}$. The left y-axis is the dynamic Kerr rotation, and the right y-axis is the spin accumulation ($\Delta M_{Cu}$) on Cu in a unit of magnetization density using a conversion factor[43] of $4\times10^{-9}$ rad m A$^{-1}$. **b**, The negative of the time-derivative of magnetization ($-dM_{FeRh}h/dt$) of FeRh, obtained from the $\Delta M_{FeRh}$ results of Fig. 2. The black squares and red circles are from ultrafast magnetization and ultrafast demagnetization, respectively, at a base temperature of 300 K and 430 K. **c,d**, The spin diffusion simulation for the (c) ultrafast-magnetization-driven and (d) ultrafast-demagnetization-driven spin current. The black lines in (c) and (d) are the z-component of the spin generation rate ($g_s$) on the FeRh, determined as $g_s = -\frac{dM_{FeRh}}{dt} \times \frac{M_z}{M_s}$, where $dM_{FeRh}h/dt$ is obtained from smooth fittings of (b), and $M_z/M_s$ is 0.13 with an external magnetic field of 0.15 T. The red, blue, and magna lines are the simulation results of the spin accumulation on the Cu surface. The number indicates the spin relaxation time ($\tau_s$) of FeRh that is used for the simulation in a unit of ps: $\tau_s$ of 0.025 ps (c) and 0.07 ps (d) for the blue line; $\tau_s$ of 0.05 ps (c) and 0.14 ps (d) for the red line; $\tau_s$ of 0.1 ps (c) and 0.28 ps (d) for the magna line. The red lines of (c) and (d) correspond to the fitting lines in (a).

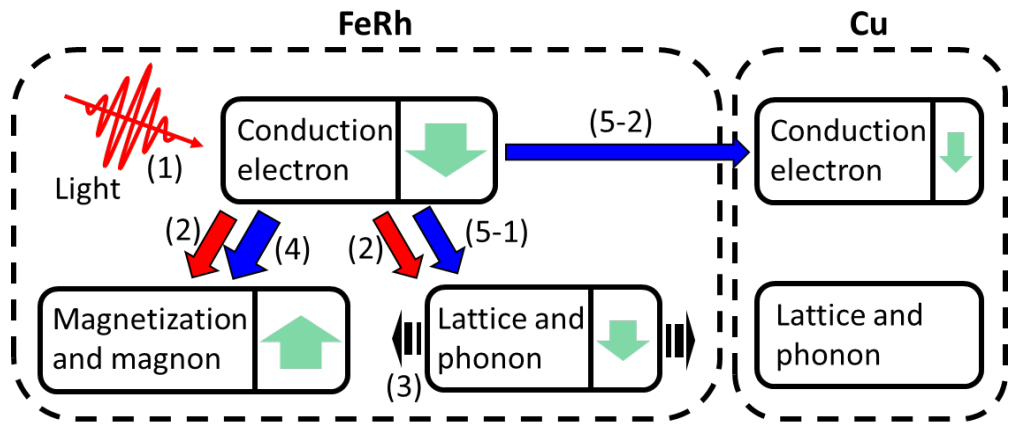

**Figure 4. Process of angular momentum transfer during phase transition of FeRh.** (1) A light energy of the pump pulse is absorbed by the electron bath of FeRh. (2) The energy flows (red arrows) from the electron bath to the magnon and phonon baths. (3) The lattice expansion determines the timescale of the phase transition dynamics. (4) During the phase transition, angular momentum (blue arrows) is suppled from the electron bath to the magnon bath. Then, the magnon bath has a positive magnetization (green arrows), and the electron bath has a negative magnetization (green arrows). The angular momentum of the electron bath of FeRh can relax to the phonon bath of FeRh (5-1) and diffuse to the electron bath of Cu (5-2). As a result of the spin diffusion from FeRh to Cu, the conduction electrons of Cu gain a negative magnetization (green arrows). The phonon bath of Cu has a negligible role for angular momentum transfer because of the long spin relaxation time of Cu.

# Supplementary Materials

# Spin current driven by ultrafast magnetization of FeRh

Kyuhwe Kang[1], Hiroki Omura[2], Oukjae Lee[3], Kyung-Jin Lee[4], Hyun-Woo Lee[5], Tomoyasu Taniyama[2], Gyung-Min Choi[1,6*]

[1]Department of Energy Science, Sungkyunkwan University, Suwon 16419, Korea

[2]Department of Physics, Nagoya University, Nagoya 464-8601, Japan

[3]Center for Spintronics, Korea Institute of Science and Technology, Seoul 02972, Korea

[4]Department of Physics, Korea Advanced Institute of Science and Technology, Daejeon 34141, Korea

[5]Department of Physics, Pohang University of Science and Technology, Pohang 37673, Korea

[6]Center for Integrated Nanostructure Physics, Institute for Basic Science, Suwon 16419, Korea

Section 1: FeRh thickness dependence on phase transition

The degree of the phase transition depends on the FeRh thickness. With FeRh thickness of 10 nm, the initial phase is not a complete AFM phase, but it is partially a FM phase with a small magnetization of 100 emu/cc at 300 K. Then, the quasi-static measurement shows that the magnetization change by the phase transition is rather a small value of 400 emu/cc (Fig. S1a). Increasing the FeRh thickness of 20 nm and 40 nm, the initial phase becomes a complete AFM phase with a negligible magnetization at 300 K. Then the magnetization change by the phase transition increases to 900 emu/cc (Fig. S1b,c). For the time-resolved measurement of ultrafast magnetization of FeRh and spin accumulation on Cu in the FeRh/Cu heterostructure, we used the FeRh thickness of 20 nm to have a complete AFM phase at 300 K.

Section 2: Optical setup for time-resolved detection

We used different optical setups depending on the experiments. Whereas the pump beam was fixed on the FeRh side of the sample, the probe beam was either on the FeRh side or Cu side depending on the experiment: probe on the FeRh side to monitor the phase transition of FeRh; probe on the Cu side to monitor the spin accumulation on Cu (Fig. S2). The probe measured the magnetization ($\Delta M_{FeRh}$) of the magnon bath ($d$ band) of FeRh or spin accumulation ($\Delta M_{Cu}$) of the conduction electron bath ($sp$ band) of Cu *via* magneto-optical Kerr effect (MOKE) using a balanced detector. The sign of MOKE is determined by the magnetization direction of the FM phase of FeRh, which is set by the external magnetic field. We applied a magnetic field of 0.15 T along the out-of-plane direction (z) using a ring magnet, whose shape allows passage of the pump and probe beams, to detect the z-components of $\Delta M_{FeRh}$ and $\Delta M_{Cu}$ using polar MOKE. If a longitudinal MOKE is preferred, one can apply the external field along the in-plane direction. Owing to the imperfection of the balanced detector, the raw signal has a small non-magnetic signal. To collect a pure magnetic signal, we simultaneously measured the Kerr rotation with +0.15 T and −0.15 T and took the difference. The probe also measured the lattice expansion of FeRh *via* the reflectivity change using a normal photodetector. For details of the reflectivity measurements, see Supplementary Section 5.

Section 3: Timescale of temperature rising

The magnitude of the temperature rising of a single FeRh layer by the pump pulse can be expressed as $\Delta T = \frac{a_{\text{FeRh}} F_{\text{pump}}}{C_{\text{FeRh}} d_{\text{FeRh}}}$, where $a_{\text{FeRh}}$ is the light absorption by FeRh, $F_{\text{pump}}$ is the incident pump fluence, $C_{\text{FeRh}}$ is the heat capacity of FeRh, and $d_{\text{FeRh}}$ is the thickness of FeRh. With $a_{\text{FeRh}}$ of 0.4, $F_{\text{pump}}$ of 7.1 J m$^{-2}$, $C_{\text{FeRh}}$ of $3.2 \times 10^6$ J m$^{-3}$ K$^{-1}$, and $d_{\text{FeRh}}$ of 20 nm, $\Delta T$ is 44 K. (The heat capacities of electron, magnon, and phonon baths of FeRh are taken from Ref. [1]. The light absorption is calculated in SI Section 5.). To estimate the dynamics of $\Delta T$ in the FeRh (20 nm)/Cu (120 nm) structure, we performed a heat transport simulation with an initial electronic heating by pump pulse (Fig. S1). Owing to a strong electron-phonon coupling of FeRh, $\Delta T$ of the phonon bath of FeRh reaches to 40 K at 1 ps. Therefore, the heating alone cannot explain the phase transition timescale. In this simulation, we assume that the electron-phonon coupling of FeRh is $10^{18}$ J m$^{-3}$ K$^{-1}$, a value of Ni [2,3], and the electron-magnon coupling is $0.5 \times 10^{18}$ J m$^{-3}$ K$^{-1}$, a value of Ni [3]. Material parameters for the thermal transport simulation are summarized in Table S1.

Section 4: Light absorption by FeRh

The light absorption by FeRh ($a_{\text{FeRh}}$) is shown as an attenuation of the Poynting vector (S), which is calculated using a transfer-matrix method with the refractive indexes of 1.73, 3+$i$5, and 0.3+$i$5 for MgO, FeRh, and Cu, respectively. When the light is incident on the FeRh side of the FeRh (20 nm)/Cu (120 nm) structure, through the MgO substrate, 55% is reflected at the MgO/FeRh interface, 44% is absorbed by the FeRh 20 nm, and 1 % is absorbed by the Cu 120 nm. The light absorption per thickness along the FeRh layer is obtained as dS/dz, where z is the position along the FeRh thickness. The dS/dz is used for the distribution of the initial electronic heating in the thermal transport simulation and for the distribution of the spin generation in the spin diffusion simulation.

Section 5: Timescale of lattice expansion

We compare timescales of the thermalization and lattice expansion from the reflectivity change ($\Delta R$) measurement. $\Delta R$ of metal films consists of temperature-induced one ($\Delta R_T$) and strain-induced one ($\Delta R_S$) [4]. A signature of the structural transition exists in $\Delta R_S$ as some of the strain change is caused by the lattice expansion during the phase transition. Let us call the lattice-expansion-driven $\Delta R_S$ as $\Delta R_L$, in distinction from the normal $\Delta R_S$, which is independent on the phase transition. A separation of $\Delta R_L$ from the others can be achieved from the magnetic-field dependence on $\Delta R$. An external magnetic field changes the degree of phase transition [5], and resulting $\Delta R_L$, of FeRh, without affecting $\Delta R_T$ and $\Delta R_S$. We collect the magnetic-field dependent part of $\Delta R$ and assign it to $\Delta R_L$. Whereas the magnetic-field independent part of $\Delta R$ shows a rapid increase at near the time zero by ultrafast heating, $\Delta R_L$ rise only after 2.5 ps. The same time delay in $\Delta M$ and $\Delta R_L$ suggests that the timescale of the magnetic ordering is limited by the timescale of the lattice expansion. We note that 2.5 ps is about half of the temporal position of 5 ps of the acoustic echo that appears in the magnetic-field independent part of $\Delta R$. The position of the acoustic echo is approximately determined as $2(d_{\text{FeRh}} - d_{\text{surf}})/v_s$, where $d_{\text{FeRh}}$ is the thickness of FeRh, $d_{\text{surf}}$ is the surface depth of initial heating, and $v_s$ is the sound velocity of ~5 km s$^{-1}$ of FeRh [1]. Due to a partial reflection at the FeRh/Cu interface, the acoustic wave makes a round trip through the FeRh layer, which takes about 5 ps. Then, the 2.5 ps delay in the lattice expansion corresponds to the time for the acoustic wave to travel through the FeRh layer.

Section 6: Materials parameters for spin diffusion simulation

For the spin generation rate ($g_s$), we used $g_s = -\frac{dM}{dt}$, where $M$ is obtained from the experimentally measured ultrafast magnetization of FeRh (Fig. 2a of the main text). Since the FeRh thickness is thicker than the light penetration depth, we assume a non-uniform $g_s$ along the FeRh thickness, and distribution of $g_s$ is determined from the distribution of the light absorption (Fig. S4). The bulk diffusivities ($D$) of FeRh and Cu are obtained from the electrical conductivities ($\sigma_e$) using the relation of $D = \frac{\sigma_e}{e^2 N_F}$. The $\sigma_e$ values of 1.3×10$^6$, 0.8×10$^6$, and 50×10$^6$ Ω$^{-1}$ m$^{-2}$ of FM FeRh, AFM FeRh, and Cu, respectively, are measured using a four-point probe method. The interfacial spin conductance ($G_s$) of the FeRh/Cu interface is obtained from the interfacial electrical conductance ($G_e$) using the relation of $G_s = \frac{G_e}{2e^2}$. Since we do not know $G_e$ value of the FeRh/Cu interface, we used $G_e$ of 2×10$^{15}$ Ω$^{-1}$ m$^{-2}$ of the permalloy/Cu and Co/Cu interfaces [6]. The diffusive spin current was estimated using $J_s = \frac{D}{N_s}\frac{\partial \mu_s}{\partial z}$ at the bulk and $J_s = G_s \Delta\mu_s$ at the interface, where $\Delta\mu_s$ is the difference in $\mu_s$ at the interface. With the thicknesses of 20 nm for FeRh and 120 nm for Cu, we found that the effect of $D$ is dominant over the effect of $G_s$. The spin relaxation time ($\tau_s$) of FeRh was determined from the fitting between the experiment and the simulation. The $\tau_s$ of Cu was determined from the reported spin diffusion length ($l_s$) of 400 nm of Cu [7]. The parameters for the spin diffusion simulation are summarized in Table S2.

**Supplementary References**

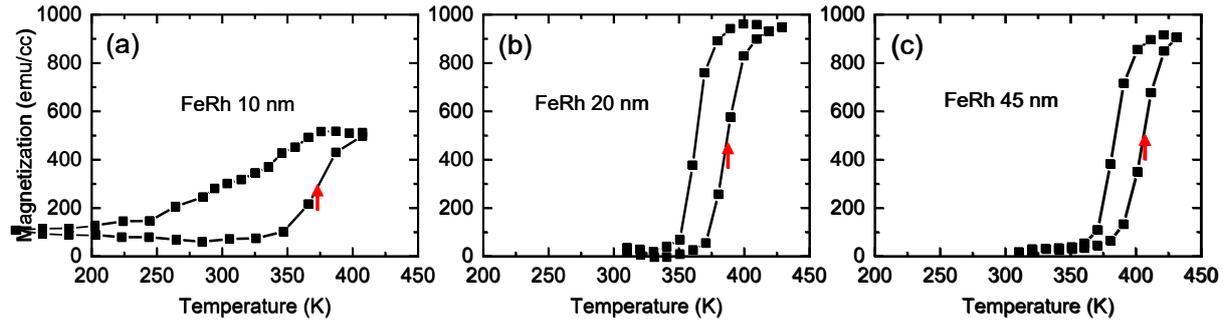

**Figure S1. Magnetization (M) versus temperature of FeRh films.** With the FeRh thickness of 10 nm (a), M of AFM phase at 300 K is ~100 emu/cc, and M of FM phase at 400 K is ~500 emu/cc. The non-zero M value of the AFM phase and rather small change of M between the AFM and FM phase suggest that the initial AFM ordering is not perfect. With the FeRh thickness of 20 nm (b) and 40 nm (c), M of AFM phase at 300 K is close to zero, and M of FM phase at 400 K is ~900 emu/cc. The negligible M value of the AFM phase and large change of M between the AFM and FM phases suggest that the initial AFM ordering is perfect. In addition, the critical temperature (red arrows) increases with the FeRh thickness: 375 K with 10 nm (a); 385 K with 20 nm (b); 405 K with 45 nm (c).

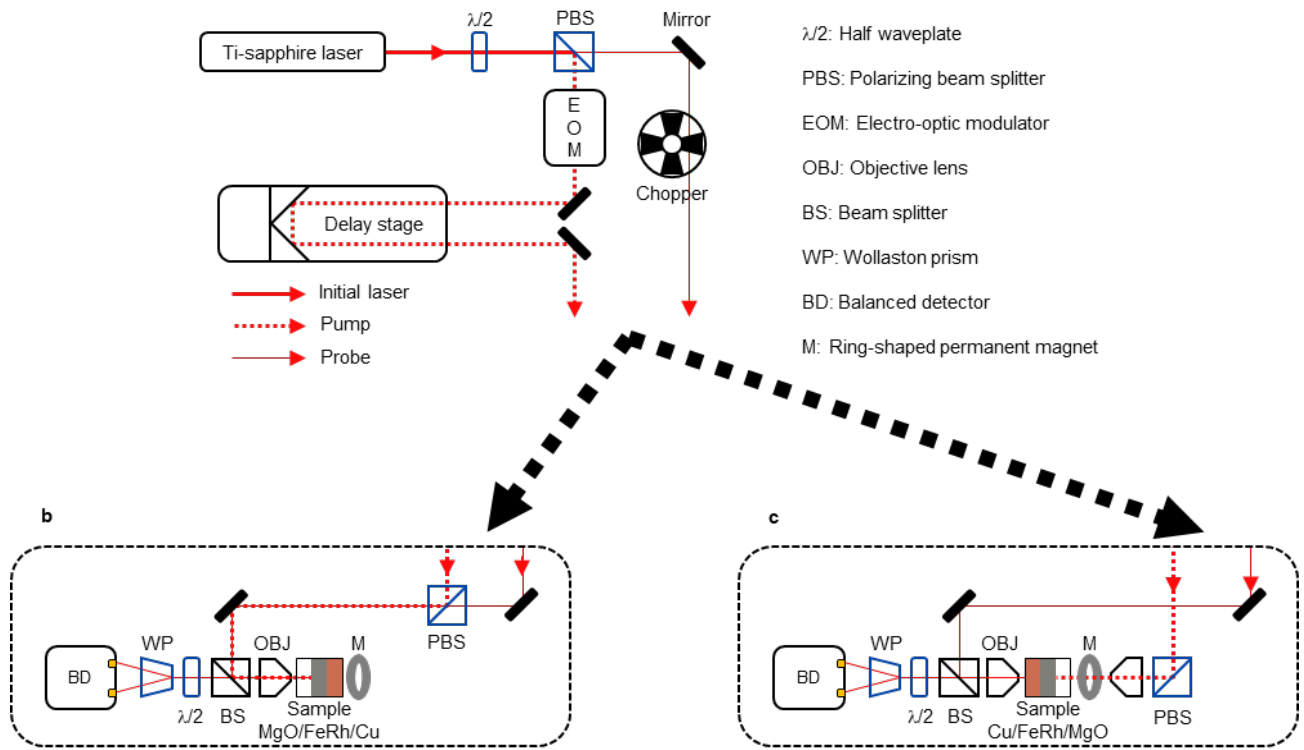

**Figure S2. Schematics of optical setup.** Ti-sapphire laser produces a pulsed laser with a wavelength of 785 nm. Polarizing beam splitter (PBS) split the laser beam into the pump (dotted line) and probe (solid line) beams. The pump and probe beams are modulated by electron-optic modulator (EOM) and chopper, respectively, at 1 MHz and 200 Hz. A delay stage controls a time delay between the pump and probe beams. Both pump and probe beams are focused on the sample surface using a 20X objective lens. The reflected probe beam passes a beam splitter (BS), halfwave plate (λ/2), and Wollaston prism (WP), then it is collected by a balanced detector (BD). To measure magnetization dynamics of FeRh, both pump and probe beams are on the FeRh side. To measure the spin accumulation on Cu, the probe is on the Cu side, and the pump is on the FeRh side. A ring magnet (M) is inserted between the sample and objective lens to align the magnetization of FeRh to the out-of-plane direction. Then, the magnetization dynamics and spin accumulation can be measured by polar MOKE.

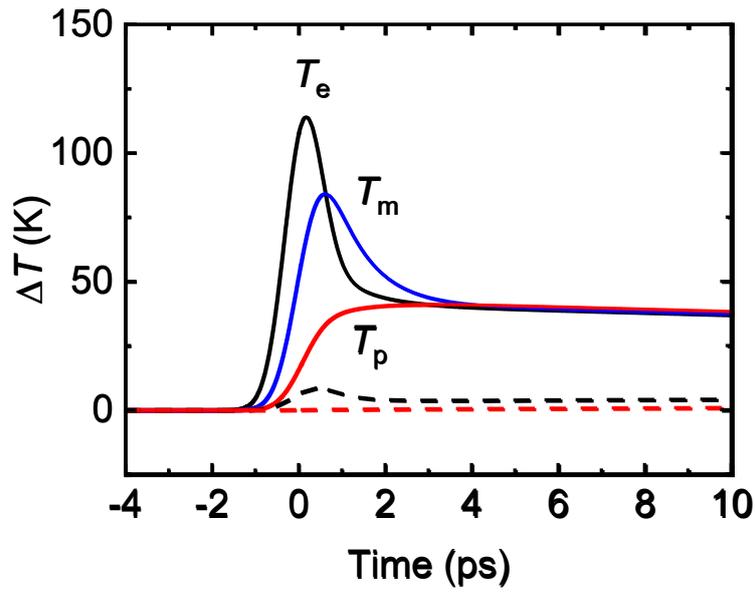

**Figure S3. Timescale of temperature rising in FeRh.** The simulation of the temperature rising ($\Delta T$) in the FeRh (20 nm)/Cu (120 nm) heterostructure with an incident pump fluence of 7.1 J m$^{-2}$. The back, red, and blue colors are for the electron ($T_e$), phonon ($T_p$), and magnon ($T_m$) temperatures, respectively. After the electronic heating by the pump pulse, there is rapid rising of $T_p$ and $T_m$ owing to the strong electron-phonon and electron-magnon couplings. The solid and dashed lines are for FeRh and Cu, respectively. A small-but-fast increase of $T_e$ of Cu is due to the fast electronic transport from FeRh to Cu during the high $T_e$ of FeRh. There is no rapid heating of Cu phonon because of the small electron-phonon coupling of Cu.

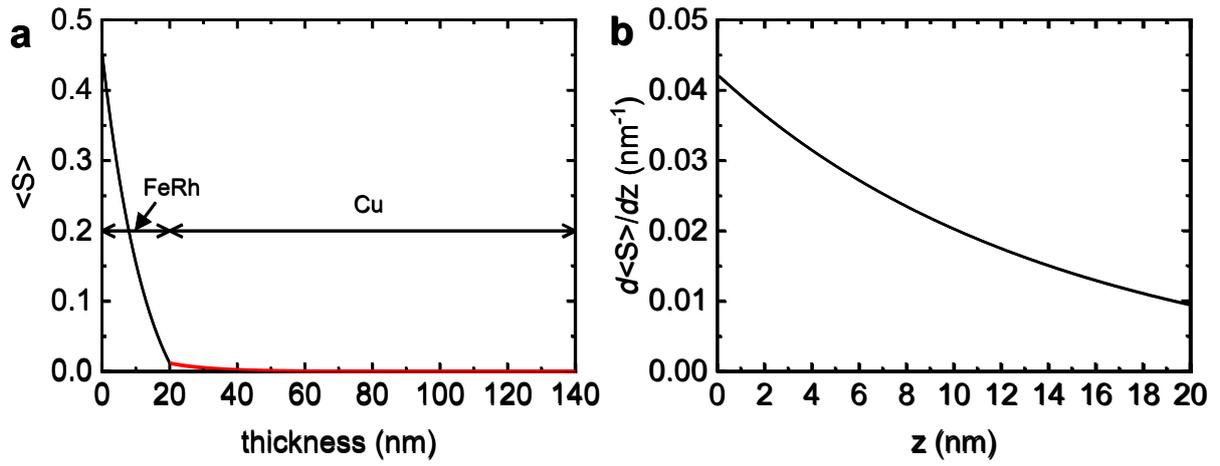

**Figure S4. Light absorption in FeRh.** (a) The attenuation of the light intensity in the MgO substrate/FeRh (20 nm)/Cu (120 nm) structure. We calculate the time-averaged Poynting vector, <S>, using a transfer-matrix method with the refractive indexes of 1.73, 3+$i$5, and 0.3+$i$5 for MgO, FeRh, and Cu, respectively. Most of the light absorption occurs in the FeRh layer (black line), and the Cu layer's contribution to the light absorption is negligible (red line). (b) The light absorption per thickness is calculated as the spatial derivative of <S> along the FeRh thickness (z). We assume that the initial electronic heating in the thermal transport simulation and the spin generation rate ($g_s$) in the spin diffusion simulation has the same distribution of (b) along the FeRh thickness.

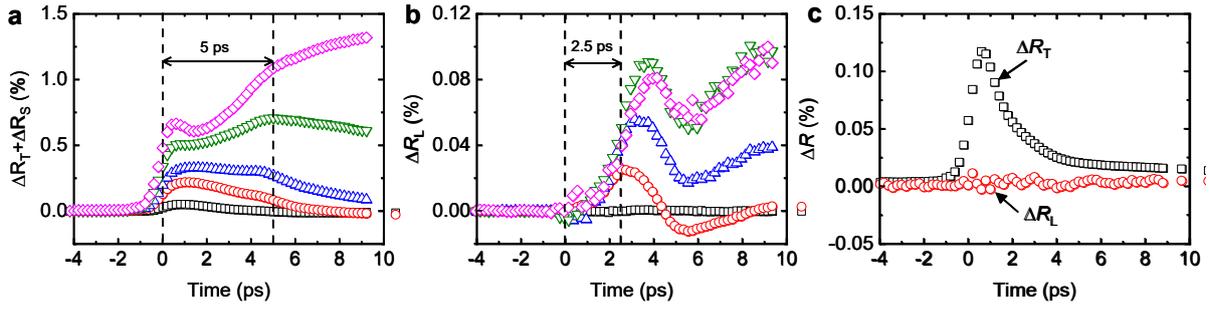

**Figure S5. Timescale of lattice expansion in FeRh. a,b,**The reflectivity change (ΔR) during the phase transition of FeRh in the FeRh [20 nm]/Cu [120 nm] sample. Comparing ΔR with and without an external magnetic field of 0.15 T, we separate the magnetic-field independent and dependent parts of ΔR. (a) The magnetic-field independent ΔR consists of the temperature-induced ($\Delta R_T$) and strain-induced ($\Delta R_S$) ones, and (b) the magnetic-field dependent ΔR comes from the lattice expansion ($\Delta R_L$) during the phase transition. The color indicates the pump fluence in a unit of J m$^{-2}$: 0.9 (black square), 2.2 (red circles), 3.5 (blue up-triangles), 5.3 (green down-triangles), and 7.1 (magna diamonds). **c,** ΔR during ultrafast demagnetization of Co in the Co [20 nm]/Cu [100 nm] sample. There is strong signal of the magnetic-field independent ΔR, which comes from the temperature-driven one ($\Delta R_T$). However, there is no magnetic-field dependent ΔR, indicating that the conventional ferromagnets does not show any structural transition during the ultrafast demagnetization.

|  | AFM FeRh | Cu |
|---|---|---|
| $C_{tot}$ ($10^6$ J m$^{-3}$ K$^{-1}$) | 3.2 | 3.4 |
| $\gamma$ (J m$^{-3}$ K$^{-2}$) | 215[a] | 97[b] |
| $C_m$ ($10^6$ J m$^{-3}$ K$^{-1}$) | 0.3[a] | 0 |
| $C_p$ ($10^6$ J m$^{-3}$ K$^{-1}$) | 2.8[a] | 3.4[b] |
| $\Lambda_e$ (W m$^{-1}$ K$^{-1}$) | 5.9[c] | 300[b] |
| $\Lambda_p$ (W m$^{-1}$ K$^{-1}$) | 5[d] | 5[d] |
| $g_{e-p}$ ($10^{17}$ J m$^{-3}$ K$^{-2}$) | 10[e] | 0.75[f] |
| $g_{e-m}$ ($10^{17}$ J m$^{-3}$ K$^{-2}$) | 5[e] |  |

**Table S1. Material properties for heat transport simulation.** $C_{tot}$ is the total heat capacity, $\gamma$ is the electronic heat capacity coefficient, $C_m$ is the magnon heat capacity, $C_p$ is the phonon heat capacity, $\Lambda_e$ is the electronic thermal conductivity, $\Lambda_p$ is the phonon thermal conductivity, $g_{e-p}$ is the electron-phonon coupling, and $g_{e-m}$ is the magnon-phonon coupling.

a. from Reference[1].

b. from Reference[8].

c. obtained from Wiedemann-Frantz law with measured electrical conductivities.

d. typical value of metals from Reference[9].

e. a value of Ni[2,3].

f. from Reference[10]

|  | FM FeRh | AFM FeRh | Cu |
|---|---|---|---|
| $N_F$ ($10^{47}$ J m$^{-3}$) | 8.25 | 3.43 | 1.6 |
| $\sigma$ ($10^6$ $\Omega^{-1}$ m$^{-1}$) | 1.3 | 0.8 | 50 |
| $D$ (nm$^2$ ps$^{-1}$) | 62 | 91 | 12200 |
| $\tau_s$ (ps) | 0.14 | 0.05 | 13 |
| $l_s$ (nm) | 2.9 | 2.1 | 400 |

**Table S2. Material properties for spin diffusion simulation.** $N_F$ is the electronic density of state at the Fermi level, $\sigma$ is the electrical conductivity, $D$ is the electrical diffusivity, $\tau_s$ is the spin relaxation time, and $l_s$ is the spin relaxation length. $N_F$ values are determined as $N_F = \frac{3\gamma}{\pi^2 k_B^2}$, where $\gamma$ is the coefficient of the electronic heat capacitance from Ref. 11, $k_B$ is the Boltzmann constant. We assume that the spin density states ($N_s$) for the spin diffusion simulation can be approximated to be a half of $N_F$. $\sigma$ values are measured using a four-point probe method. $D$ values are obtained from $\sigma$ values using the relation of $D = \frac{\sigma}{e^2 N_F}$, where $e$ is the elementary charge. $\tau_s$ values of FeRh are determined by comparing the spin accumulation experiment and spin diffusion simulation (see Methods). $\tau_s$ values of Cu are obtained from $l_s$ of 400 nm [7] using the relation of $l_s = \sqrt{D\tau_s}$.